# The Rhetorical Structure of Science? A Multidisciplinary Analysis of Article Headings[1]

Mike Thelwall, Statistical Cybermetrics Research Group, University of Wolverhampton, UK.

An effective structure helps an article to convey its core message. The optimal structure depends on the information to be conveyed and the expectations of the audience. In the current increasingly interdisciplinary era, structural norms can be confusing to the authors, reviewers and audiences of scientific articles. Despite this, no prior study has attempted to assess variations in the structure of academic papers across all disciplines. This article reports on the headings commonly used by over 1 million research articles from the PubMed Central Open Access collection, spanning 22 broad categories covering all academia and 172 out of 176 narrow categories. The results suggest that no headings are close to ubiquitous in any broad field and that there are substantial differences in the extent to which most headings are used. In the humanities, headings may be avoided altogether. Researchers should therefore be aware of unfamiliar structures that are nevertheless legitimate when reading, writing and reviewing articles.

## 1 Introduction

Writing academic articles is a difficult task, especially for a multidisciplinary audience. Authors must decide on the content to include – and which details are inessential enough to omit – and on an appropriate rhetorical structure (e.g., Kanoksilapatham, 2005; Tessuto, 2015). The linguistic task of writing an academic paper consists in constructing a series of "moves" to persuade the reader of the key message (Swales, 1990, 2004). Each move performs a separate task. For example, the introductory part of an article may contain moves that argue for the importance of the topic, review prior research, identify a gap, and explain how the gap is addressed in the new paper. Moves or groups of moves can be expected to be flagged by appropriate headings.

Some journals, particularly in medicine, and some publishers, such as Emerald, mandate the use of structured abstracts with a pre-defined set of headings (Hartley, 2014). These may inform authors about the type of content to include in the remainder of the article and may influence their choice of structure. Journals may also guide the choice of rhetorical structure by requiring or recommending headings. The International Committee of Medical Journal Editors (ICMJE), for example, recommends IMRaD (Introduction, Methods, Results, and Discussion) within its guidelines (ECMJE, 2004). Even in the absence of this advice, authors, editors and referees need to ensure that the structure of an article is appropriate for its content. This may be confusing in the current era of interdisciplinary research and large general journals, such as PLoS ONE, since even experienced authors may be unsure about which structure will be most effective at communicating with their intended audience. These are important issues for science because researchers tend to be specialists in their own narrow fields but are called to review articles that only partially

---

overlap with their expertise or main interests (e.g., Lee, Sugimoto, Zhang, & Cronin, 2013). One of the goals of a review is to help the authors present their argument clearly (Taylor & Francis, 2018). Whilst reviewers seem to be harshest within their topic focus (Boudreau, Guinan, Lakhani, & Riedl, 2016; Gallo, Sullivan, & Glisson, 2016), they may make misjudged recommendations for structural changes outside of their topic focus, compromising the clarity of the work. The following APA Publications and Communications Board Task Force Report warning to those reviewing the work of qualitative psychologists illustrates that structural misunderstanding is an important issue (Levitt, Bamberg, Creswell, Frost, Josselson, & Suárez-Orozco, 2018).

> qualitative researchers often combine Results and Discussion sections, as they may see both as intertwined and therefore not possible to separate a given finding from its interpreted meaning within the broader frame of the analysis. Also, they may use headings that reflect the values in their tradition (such as "Findings" instead of "Results") and omit ones that do not.

From prior small-scale studies, it is known that article structures are not uniform across science. Based on an analysis of several small corpora of scientific documents, Teufel (1999) argues that scientific texts can be split into seven "argumentative zones": Background; Topic; Related work; Purpose/problem; Solution/method; Result; Conclusion/claim (Teufel, 1999: p. 108). These zones were sometimes, but not always, flagged by headings, could be separate or mixed sections, and could be arranged in different orders based on discipline or national culture. For example, she found that cardiology papers (n=103 articles) tended to follow standard heading names and structure, with at least 92% using Introduction, Methods, Results and Discussion section names. In contrast, computational linguistics papers (n=80 papers) did not follow a standard structure: although 79% had an Introduction section and 59% had a Conclusion/s section, the most common other standard section was shared by only 16% (Discussion) (Teufel, 1999: p. 82). Another study with similar goals attempted to classify text within into eleven core functions (Background, Hypothesis, Goal, Motivation, Object, Method, Model, Experiment, Observation, Result, Conclusion) for 265 biochemistry and chemistry articles (Liakata, Saha, Dobnik, Batchelor, & Rebholz-Schuhmann, 2012). There are many small scale qualitative studies of the rhetorical structure of academic articles, which give evidence that articles do have recognisable structures but with great variety between them (e.g., Kanoksilapatham, 2005; Tessuto, 2015). For example, an analysis of 67 articles in five engineering fields found very few overall similarities, although most had introductory and concluding sections (Maswana, Kanamaru, & Tajino, 2015).

Despite the insights given by prior research, as briefly reviewed above, there is no systematic body of evidence about differences in style between and within disciplines. This is necessary to inform authors as well as reviewers evaluating the communicational effectiveness of a structure. The following questions drive the investigation.

1. What are the main disciplinary differences in the headings used to organise research articles?
2. Are there universally accepted sets of headings for research articles in any disciplines?

# 2 Methods

The research design was to obtain a large multidisciplinary collection of full text articles, extract their headings and summarise the frequency of these headings by broad and narrow field category. The main initial decision was the choice of full text collection.

At a computational level, previous work has shown that it is possible to extract article sections from HTML or PDF versions of articles with a reasonable degree of accuracy (Ding, Liu, Guo, & Cronin, 2013; Ramakrishnan, Patnia, Hovy, & Burns, 2012; Ronzano & Saggion, 2015) and argumentative zones within sections from full text collections (Teufel, 1999; Teufel & Moens, 2002). This approach was rejected on the basis that it was unnecessary given the existence of a large multidisciplinary collection of full text journal articles in structured form.

The PMC (PubMed Central) Open Access Subset (PMC, 2018a) from the US National Institutes of Health was used as the raw data, as downloaded 27-28 October 2018. This is a collection of over 2 million full text copies of articles that are indexed in PMC (which indexes 3189 journals: PMC, 2018b) available for academic research. It is the largest recent systematic multidisciplinary collection of open access full text articles. It is a voluntary collection that includes large number of articles from open access journals, including PLOS ONE, as well as open access articles from mixed journals. It is voluntary for journals and authors and probably excludes most Web of Science and Scopus journals. It focuses on "biomedical and life sciences" and started in 2000 (PMC, 2018) with the earliest dated articles from 1998.

PMC articles in XML follow the National Information Standards Organization (NISO) Journal Archiving and interchange Tag Sets (JATS) Version 1.1 (NCBI, 2015a) or an earlier version. This includes tags for metadata, section structures and section titles. Articles are also assigned a type, such as: research-article, review-article, systematic-review, brief-report, case-report, letter, rapid-communication, editorial, article-commentary, reply, news, in-brief, book-review, other, abstract, calendar, correction, retraction, obituary, introduction, meeting-report, addendum. Only documents declared to be the first type (research-article) were retained. The advisory description for these is, "Article reporting on primary research (The related value "review-article" describes a literature review, research summary, or state-of-the-art article.)" (NCBI, 2015ab). This description therefore encompasses standard non-review articles and should largely exclude shorter form and non-reviewed contributions. The research-article type was the most common in 90% of the PMC journals (7702 out of 8501).

Articles in PMC are given MeSH (Medical Subject Headings) terms for indexing but not a subject classification. They were classified by subject using the ScienceMetrix journal list of 22 broad and 176 narrow field categories (Archambault, Beauchesne, & Caruso, 2011; http://science-metrix.com/?q=en/classification). Although this is less accurate than article-level clustering, it is probably better than the Scopus and Web of Science categories (Klavans & Boyack, 2017). Article-level clustering was not used because the dataset is not balanced by discipline and it is unclear whether clustering would work well. In any case, it would likely be much more fine-grained for biomedicine than for other fields. The 100 journals not in the ScienceMetrix list (either absent of with a different title) with most articles were manually assigned an appropriate category. This included 21 "Frontiers in" journals as well as the high profile open access journals Nature Reports, SpringerPlus (see Appendix Table SM1).

Articles structures were detected from their XML section headings, either using their internal standardised names (e.g., <sec id="Sec2" sec-type="methods">) or the contents of the title tag immediately following the section tag (e.g., <title>methods</title>), if any. Many sections did not have a title but were used for organising the text. Common title variations were standardised (e.g., method -> methods; see Appendix, Table A1) and initial numbers were removed (e.g., "1. Introduction" -> "Introduction") as well as tags (e.g., <italic>). The sec-type tag attribute was allowed to override the official title of any section to reduce the impact of minor heading variations (e.g., *Introductory remarks* rather than *Introduction*).

The number of times each section title was used was tallied by journal category, after deleting multiple instances of the same title within single articles. As a convenient cut-off, sections occurring at least 100,000 times overall were singled out for discussion, as well as the most common other type of section in each ScienceMetrix main category.

## 3   Results

The results show substantial variations between disciplines and a lack of ubiquitous article headings (and hence structures) within four groups of broad disciplines (Figure 1, see also Appendix Table A2). This also extends to broad disciplines (Table 1).

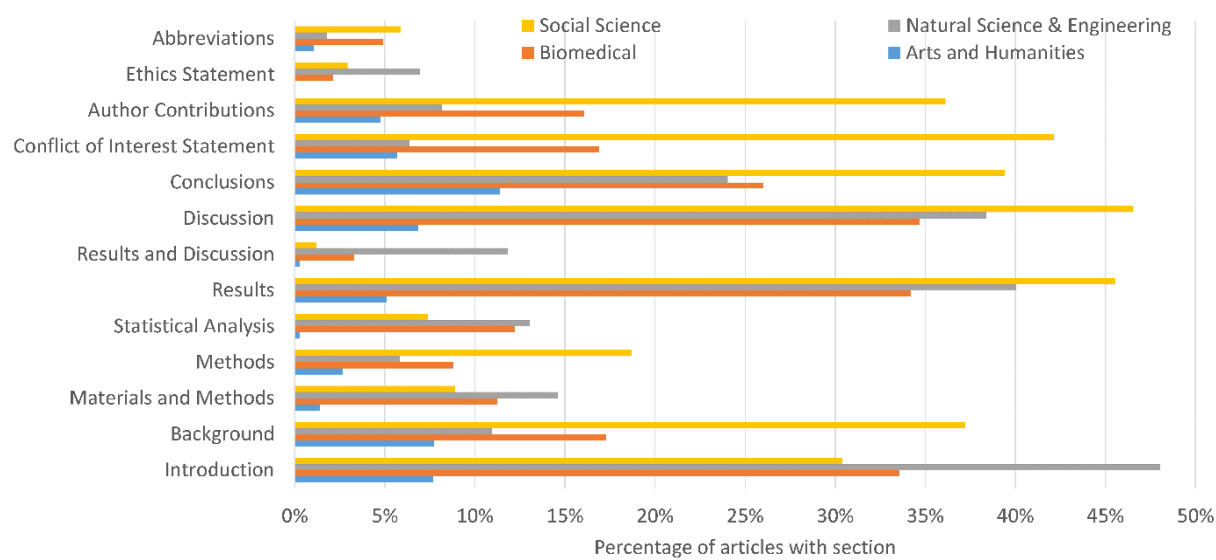

Figure 1. Main sections occurring at least 100,000 times (8222 journals; 1584674 research articles), as a percentage of articles in the category: Arts and Humanities (General Arts, Humanities & Social Sciences; Historical Studies; Philosophy & Theology; Visual & Performing Arts), Biomedical (Biomedical Research; Clinical Medicine); Natural Science and Engineering (Agriculture, Fisheries & Forestry; Biology; Built Environment & Design; Chemistry; Earth & Environmental Sciences; Enabling & Strategic Technologies; Engineering; General Science & Technology; Information & Communication Technologies; Mathematics & Statistics; Physics & Astronomy); Social Science (Communication & Textual Studies; Economics & Business; Psychology & Cognitive Sciences; Public Health & Health Services; Social Sciences).

*Introduction*: Whilst in some disciplines articles usually started with explicitly labelled introductions (e.g., Physics & Astronomy: 86%), in others they were rare (e.g., Historical Studies: 8%). Articles could also have an implicit introduction section in the form of an unlabelled initial set of paragraphs. An example of a non-standard article structure within the relatively homogenous Physics & Astronomy broad category is the paper, "Correlation Inequalities for the Quantum XY Model" in *Journal of Statistical Physics*, which started with a section labelled: *Introduction and Results*. The remaining section headings were all unique, starting with *Infinite Volume Limit of Correlation Functions*.

*Background*: Explicitly labelled background sections were only the norm in Public Health & Health Services (60%) and were rare in many (e.g., Chemistry: 7%). A background section could involve a literature review, theory or technical details necessary to understand the methods.

*Methods*: Methods sections were commonly also labelled Materials and Methods. These were surprisingly far from ubiquitous, with Earth & Environmental Sciences (Materials and Methods: 40%; Methods: 11%) being the main users. These sections were absent from three arts, humanities and social science categories. Statistical analysis sections were also rare but least uncommon in Clinical Medicine (19%) and General Science and Technology (19%). A previous analysis of a mid-2015 version of this dataset found 35% of articles to have at least one main section with a title containing the word stem "method" (Small, 2018). This higher figure is presumably due to including many names for methods sections not included in Table 1 (e.g., Statistical Methods, Patients and Methods, Research Design and Methods).

*Results and discussion*: These sections were often combined but were only in most articles within three broad categories: Agriculture, Fisheries & Forestry (55%), Public Health & Health Services (Results: 54%) and Engineering (Results: 52%). Combining results with discussion into a single section was most common in Chemistry (42%), Physics and Astronomy (35%) and Earth and Environmental Sciences (28%). Five broad categories did not use a combined results and discussion section.

*Conclusions*: Only a quarter (27%) of articles had an explicitly labelled conclusion section, although they were in a majority in eight broad categories. They were most common in Information & Communication Technologies (67%) and occurred to some extent in all categories. Perhaps surprisingly, only 26% of Biomedical Research articles had a labelled conclusion. This could be due to following IMRaD guidelines for section headings.

*Discipline-specific sections* (end 3 columns of Table 2): Except for Historical Studies, which rarely used section titles but published essays, and Physics & Astronomy, all broad disciplines included an average of two non-standard section names within articles. Non-standard section names were most common in Information & Communication Technologies (5.7 non-standard section names per article). Some common non-standard section names were generic, such as *Experimental section*, whereas others were discipline-specific, such as *Animals* and *Plant material*.

Table 1. Main sections occurring at least 100,000 times by broad Science-Metrix category (8222 journals; 1584674 research articles), as a percentage of articles in the category.

| Broad category | Journals | Articles | Intro-duction % | Back-ground % | Materials & methods % | Methods % | Statistical analysis % | Results % | Results and discussion % | Discussion % | Conclusion % |
|---|---|---|---|---|---|---|---|---|---|---|---|
| Ag., Fish. & Forestry | 105 | 15082 | 44 | 17 | 31 | 11 | 17 | 55 | 8 | 55 | 41 |
| Biology | 179 | 36014 | 53 | 20 | 16 | 8 | 9 | 48 | 9 | 47 | 29 |
| Biomedical Research | 435 | 212212 | 42 | 16 | 16 | 8 | 11 | 39 | 8 | 39 | 26 |
| Built Env. & Design | 17 | 66 | 61 | 3 | 6 | 14 | 5 | 26 | 5 | 23 | 38 |
| Chemistry | 147 | 27812 | 76 | 7 | 14 | 4 | 1 | 15 | 42 | 12 | 60 |
| Clinical Medicine | 1222 | 345324 | 35 | 22 | 14 | 13 | 19 | 45 | 2 | 46 | 30 |
| Comm. & Textual Stud. | 15 | 26 | 8 | 8 | 0 | 8 | 0 | 12 | 0 | 4 | 4 |
| Earth & Environ. Sci. | 91 | 1816 | 67 | 14 | 40 | 11 | 1 | 44 | 28 | 37 | 52 |
| Economics & Business | 77 | 263 | 56 | 6 | 3 | 13 | 1 | 28 | 1 | 21 | 37 |
| Enabling & Strategic Tech. | 145 | 62215 | 33 | 50 | 14 | 17 | 3 | 40 | 30 | 37 | 58 |
| Engineering | 102 | 2485 | 40 | 46 | 18 | 25 | 5 | 52 | 15 | 49 | 57 |
| Gen. Arts Hum. & Soc. Sci. | 13 | 87 | 18 | 3 | 0 | 0 | 2 | 1 | 0 | 1 | 8 |
| Gen. Science & Tech. | 13 | 310468 | 50 | 2 | 15 | 3 | 19 | 45 | 6 | 43 | 10 |
| Historical Studies | 45 | 1673 | 8 | 0 | 2 | 0 | 0 | 2 | 0 | 3 | 11 |
| Info. & Comm. Tech. | 84 | 6352 | 24 | 51 | 8 | 21 | 10 | 44 | 1 | 44 | 67 |
| Math. & Statistics | 65 | 523 | 74 | 2 | 0 | 9 | 0 | 9 | 0 | 45 | 13 |
| Philosophy & Theology | 36 | 1221 | 35 | 43 | 7 | 17 | 2 | 30 | 1 | 39 | 54 |
| Physics & Astronomy | 89 | 11961 | 86 | 0 | 43 | 3 | 1 | 48 | 35 | 47 | 63 |
| Psych. & Cognitive Sci. | 216 | 20704 | 53 | 3 | 7 | 10 | 9 | 43 | 1 | 39 | 23 |
| Pub. Health & H. Serv. | 250 | 43262 | 21 | 60 | 11 | 26 | 9 | 54 | 1 | 58 | 52 |
| Social Sciences | 195 | 1448 | 53 | 6 | 9 | 17 | 1 | 32 | 2 | 29 | 27 |
| Visual & Perform. Arts | 4 | 6 | 0 | 0 | 0 | 0 | 0 | 0 | 0 | 0 | 17 |
| No category (NA) | 4677 | 483654 | 36 | 8 | 13 | 8 | 13 | 26 | 7 | 28 | 25 |
| **Total without NA** | **3545** | **1101020** | **43** | **17** | **15** | **9** | **15** | **44** | **8** | **43** | **27** |
| **Maximum** | **4677** | **483654** | **86** | **60** | **43** | **26** | **19** | **55** | **42** | **58** | **67** |
| **Minimum** | **4** | **6** | **0** | **0** | **0** | **0** | **0** | **0** | **0** | **0** | **4** |

There are substantial variations between and within broad disciplines in the extent that common type of supporting information is reported in standard named sections within the main body of an article (Table 2). A relatively rare non-standard section *Pre-publication history*, was also reported in a substantial minority of some broad categories, including Information & Communication Technologies (27%). This is presumably due to publisher strategies in these areas as well as disciplinary cultures.

Table 2. Supporting sections occurring at least 100,000 times by broad Science-Metrix category, the ratio of other sections per article, and the most common other section for each category (8222 journals; 1584674 research articles).

| Broad category | Conflict of interest % | Author Contri-butions % | Ethics State-ment % | Abbrev-iations % | Other sections per article | Main other section | % |
|---|---|---|---|---|---|---|---|
| Ag., Fish. & Forestry | 11 | 12 | 1 | 3 | 3.4 | Animals | 4 |
| Biology | 26 | 32 | 0 | 3 | 5.1 | Plant material | 4 |
| Biomedical Research | 15 | 17 | 3 | 4 | 5.5 | Funding | 5 |
| Built Env. & Design | 3 | 0 | 0 | 0 | 3.6 | Methodology | 9 |
| Chemistry | 4 | 4 | 0 | 1 | 2.5 | Experimental section | 15 |
| Clinical Medicine | 21 | 19 | 3 | 7 | 3.9 | Pre-publication history | 6 |
| Comm. & Textual Stud. | 0 | 0 | 0 | 0 | 4.2 | Participants | 8 |
| Earth & Environ. Sci. | 8 | 8 | 0 | 1 | 2.1 | Experimental section | 3 |
| Economics & Business | 2 | 0 | 0 | 0 | 4.3 | Data | 11 |
| Enabling & Strategic Tech. | 23 | 34 | 1 | 9 | 3.9 | Availability & requirements | 3 |
| Engineering | 28 | 29 | 0 | 8 | 3.0 | Limitations | 2 |
| Gen. Arts Hum. & Soc. Sci. | 0 | 0 | 0 | 0 | 4.2 | General | 8 |
| Gen. Science & Tech. | 0 | 0 | 11 | 0 | 4.8 | Participants | 3 |
| Historical Studies | 0 | 0 | 0 | 0 | 1.4 | Funding | 1 |
| Info. & Comm. Tech. | 28 | 27 | 0 | 6 | 5.7 | Pre-publication history | 27 |
| Math. & Statistics | 1 | 0 | 0 | 0 | 4.2 | Simulation study | 12 |
| Philosophy & Theology | 29 | 24 | 0 | 4 | 3.4 | Pre-publication history | 20 |
| Physics & Astronomy | 1 | 0 | 0 | 0 | 1.2 | Experimental section | 37 |
| Psych. & Cognitive Sci. | 56 | 40 | 10 | 0 | 5.4 | Participants | 33 |
| Pub. Health & H. Serv. | 39 | 37 | 0 | 10 | 3.1 | Pre-publication history | 27 |
| Social Sciences | 0 | 0 | 0 | 0 | 3.7 | Data and methods | 7 |
| Visual & Perform. Arts | 0 | 0 | 0 | 0 | 4.0 | Data | 33 |
| No category (NA) | 9 | 6 | 1 | 1 | 3.3 | Related literature | 5 |
| **Total without NA** | **15** | **15** | **5** | **4** | **4.4** | | |
| **Maximum** | **56** | **40** | **11** | **10** | **5.7** | | |
| **Minimum** | **0** | **0** | **0** | **0** | **1.2** | | |

Most (172 out of 176) of the **narrow** Science-Metrix categories had articles in the dataset, but many had small numbers. For narrow categories with at least 40 different journals and 400 articles (to ensure reasonable coverage to draw conclusions from), no section heading was close to ubiquitous. The closest was Background (72% of Public Health research articles in Table 3) and all main sections were absent from 76% of articles in at least one category (e.g., *Discussion* was in a least 24% of all articles in all categories). In the data from narrow categories with fewer articles or journals, the largest narrow category with close to ubiquity for any section was Food Science, with 97% of its 3898 articles from 17 journals containing an Introduction. No other heading was used by more than 90% of articles in any narrow category with at least 20 articles.

Table 3. Main sections occurring at least 100,000 times by **narrow** Science-Metrix category (qualification: at least 40 journals and 400 articles), as a percentage of articles in the category.

| Narrow category | Journals | Articles | Intro-duction % | Back-ground % | Materials & methods % | Methods % | Statistical analysis % | Results % | Results and Discussion % | Dis-cussion % | Concl-usion % |
|---|---|---|---|---|---|---|---|---|---|---|---|
| Biochemistry & Molecular Biology | 82 | 39192 | 58 | 8 | 37 | 8 | 10 | 46 | 10 | 46 | 35 |
| Biotechnology | 42 | 9000 | 39 | 47 | 27 | 20 | 5 | 47 | 29 | 44 | 51 |
| Cardiovascular System & Hematology | 103 | 9839 | 30 | 31 | 11 | 19 | 29 | 43 | 0 | 46 | 40 |
| Developmental Biology | 68 | 69468 | 39 | 7 | 6 | 2 | 7 | 27 | 6 | 27 | 9 |
| Endocrinology & Metabolism | 58 | 13136 | 34 | 4 | 16 | 10 | 9 | 41 | 1 | 34 | 20 |
| Experimental Psychology | 61 | 6459 | 45 | 1 | 10 | 8 | 9 | 35 | 1 | 30 | 19 |
| General & Internal Medicine | 57 | 37956 | 33 | 19 | 8 | 11 | 18 | 22 | 1 | 38 | 24 |
| Immunology | 79 | 39292 | 19 | 7 | 9 | 4 | 12 | 32 | 5 | 24 | 12 |
| Medicinal & Biomolecular Chemistry | 42 | 5357 | 55 | 27 | 19 | 8 | 3 | 26 | 52 | 24 | 60 |
| Microbiology | 74 | 27261 | 32 | 31 | 8 | 11 | 18 | 46 | 9 | 45 | 36 |
| Neurology & Neurosurgery | 147 | 33717 | 33 | 20 | 13 | 12 | 18 | 39 | 2 | 37 | 29 |
| Nuclear Medicine & Medical Imaging | 53 | 4836 | 57 | 28 | 35 | 22 | 9 | 65 | 3 | 61 | 53 |
| Nursing | 41 | 1343 | 14 | 33 | 6 | 18 | 3 | 30 | 2 | 34 | 34 |
| Oncology & Carcinogenesis | 114 | 81528 | 42 | 18 | 14 | 8 | 27 | 57 | 2 | 57 | 24 |
| Pharmacology & Pharmacy | 76 | 9509 | 49 | 4 | 12 | 8 | 28 | 40 | 14 | 38 | 29 |
| Plant Biology & Botany | 49 | 16931 | 53 | 17 | 10 | 6 | 15 | 52 | 10 | 51 | 31 |
| Psychiatry | 51 | 6998 | 32 | 43 | 17 | 24 | 12 | 54 | 1 | 58 | 45 |
| Public Health | 66 | 19572 | 19 | 72 | 11 | 28 | 11 | 61 | 2 | 66 | 60 |
| Social Psychology | 46 | 554 | 61 | 2 | 13 | 30 | 2 | 46 | 5 | 42 | 32 |
| Surgery | 46 | 3320 | 42 | 25 | 28 | 22 | 9 | 63 | 1 | 68 | 44 |
| **Maximum in this set** | **147** | **81528** | **61** | **72** | **37** | **30** | **29** | **65** | **52** | **68** | **60** |
| **Minimum in this set** | **41** | **554** | **14** | **1** | **6** | **2** | **2** | **22** | **0** | **24** | **9** |

## 4   Discussion

The results are limited by the methodological assumption that all XML records contain section names consistently in one of the two standard formats checked. Whilst no exceptions have been found, there were too many journals to check individually. The largest limitation is that the collection comprises only open access articles within the PMC Open Access subset. For the category-specific analyses, the journals also need to be registered within the Science-Metrix classification scheme (and declared extensions). Thus, whilst all journals covered should be reputable, the coverage of science is uneven and the collection is a biased subset of all categories. Another important limitation is that the dataset may include short contributions that do not need headings, even though only documents of declared type research-article were processed. The percentages reported in the tables above should therefore be taken as applicable to a subset of articles within the category rather than category-wide estimates. A related issue is that articles may not report primary research and so may avoid using standard section headings even within a specialism that requires or encourages them. Similarly, there may be specialisms that have strong section naming norms for types of article but include a mix of article types (e.g., qualitative, quantitative).

The results agree with prior smaller scale studies to the extent that they show differences between and within disciplines in the typical structures of articles. The tables above extend these results by showing, for the first time, that there are no broad disciplines with homogenous publishing structures. The same is also true for the 20 narrow categories that were extensively enough covered in the dataset for reasonable conclusions to be drawn. In contrast to the previous finding that 92% of 103 cardiology articles followed IMRaD (Teufel, 1999), however, the current study found much lower levels of ubiquity within all broad categories and the 20 largest narrow categories.

Since this study found no evidence that headings are ubiquitous in any broad category or any large enough narrow category (meeting the thresholds discussed above), it seems likely that most researchers will have read articles with some structural variations, even if they operate within a narrow specialism. This may not be true for all narrow specialisms, however, especially if they are based around journals with strong structural guidelines, such as for IMRaD.

If the thresholds for analysing the narrow categories are relaxed to include all those with at least 100 articles (n=101), any number of journals, and headings occurring at least 100 times then evidence of greater uniformity in the common headings can be found. In addition to the standard headings being the norm in at least one narrow category (Introduction: up to 94%; Background: 72%; Methods: 49%; Materials and Methods: 65%; Statistical Analysis: 50%; Results: 83%; Results and Discussion: 70%; Discussion: 82%; Conclusions: 80%), a few other less common headings occurred in at least a quarter of articles in at least one narrow category (Participants: up to 34%; Experimental: 29%; Experimental Section: 38%; Research Design and Methods: 30%; Taxonomy: 27%; Data and Methods: 25%). From this list it is clear than none of the rarer (non-administrative) headings ignored in the main analysis were close to ubiquitous within a narrow category with at least 100 articles.

The 50 most common headings not reported in the tables above illustrate the variety of academia and the existence of headings that are relatively common within specialist fields (relatively generic, non-administrative headings are in italic): pre-publication history, participants, animals, *statistics*, cell culture, *limitations*, *related literature*, funding, subjects, *study design*, *study population*, *experimental*, patients, western blot analysis, immunohistochemistry, *experimental section*, *data collection*, western blotting, *materials*, crystal data, financial support and sponsorship, *procedure*, flow cytometry, *measures*, *statistical methods*, *analysis*, mice, case report, case presentation, reagents, coi-statement, ethics, ethical considerations, patients and methods, western blot, immunofluorescence, antibodies, ethical approval, supplementary data, *strengths and limitations*, *summary*, consent, *design*, *experimental design*, immunoblotting, chemicals, *objectives*, *research design and methods*, disclosures, list of abbreviations.

# 5 Conclusions

The results show, apparently for the first time, that the headings used in scientific articles vary throughout academia, with none being close to ubiquitous in any broad category. Whilst this conclusion might already be part of the tacit knowledge of experienced multidisciplinary researchers, clear statistics are useful to guide less experienced researchers and those with a narrower monodisciplinary focus. The following specific recommendations follow.

- Authors should consider disciplinary norms for the audiences of their article as well as the content to be communicated before designing a structure for their paper. If publishing in a different discipline, they should make themselves aware of the norms of that discipline and the extent to which norm violation is tolerated.
- Reviewers should be aware that articles following unfamiliar structures are not necessarily in need of correction: this judgement should consider disciplinary norms and article content as well as journal guidelines.
- Readers should not judge an article to be poor if it follows a non-standard structure, especially if it from outside their home discipline(s) or if the home discipline tolerates structural variations.

Finally, the structural heterogeneity of academia contrasts with the proscriptive nature of IMRaD. Its support by many editors within medicine presumably indicates that it is useful in this area. In parallel with evidence of the communicational efficacy of structured abstracts (Hartley, 2014), it may be worth considering whether the current structural variety is optimal from a communication perspective, or whether the adoption of additional discipline-sensitive guidelines may be helpful.

# 6 Supplementary material

Counts of article headings for each broad and narrow category and journal are in the online supplementary material doi:10.6084/m9.figshare.7445603. The supplementary material in the publisher website contains tables of results for all document types (not just research articles) and the journals added to the Science-Metrix classification scheme.

# 8 Appendix

Table A1. Rules used to standardise heading names

| Original heading | Converted to |
|---|---|
| intro | introduction |
| materials\|methods | materials and methods |
| material and methods | materials and methods |
| methods and materials | materials and methods |
| method | methods |
| conclusion | conclusions |
| author contribution | author contributions |
| authors' contributions | author contributions |
| authors' contribution | author contributions |
| conflict of interest | conflict of interest statement |
| conflict of interests | conflict of interest statement |
| conflicts of interest | conflict of interest statement |
| competing interest | conflict of interest statement |
| competing interests | conflict of interest statement |

Table A2. Main sections occurring at least 100,000 times by groups of broad Science-Metrix categories, as a percentage of articles in the category.

| Section | Arts and Humanities | Biomedical | Natural Science & Engineering | Social Science | No category (NA) | Total without NA |
|---|---|---|---|---|---|---|
| Introduction | 8% | 34% | 48% | 30% | 29% | 39% |
| Background | 8% | 17% | 11% | 37% | 7% | 16% |
| Materials and Methods | 1% | 11% | 15% | 9% | 8% | 12% |
| Methods | 3% | 9% | 6% | 19% | 6% | 8% |
| Statistical Analysis | 0% | 12% | 13% | 7% | 8% | 12% |
| Results | 5% | 34% | 40% | 46% | 17% | 37% |
| Results and Discussion | 0% | 3% | 12% | 1% | 4% | 6% |
| Discussion | 7% | 35% | 38% | 47% | 20% | 37% |
| Conclusions | 11% | 26% | 24% | 39% | 20% | 26% |
| Conflict of Interest Statement | 6% | 17% | 6% | 42% | 8% | 14% |
| Author Contributions | 5% | 16% | 8% | 36% | 5% | 14% |
| Ethics Statement | 0% | 2% | 7% | 3% | 1% | 4% |
| Abbreviations | 1% | 5% | 2% | 6% | 1% | 4% |
| **Journals** | **100** | **1712** | **1079** | **758** | **4852** | **3649** |
| **Articles** | **8002** | **754076** | **518412** | **74602** | **822864** | **1355092** |